\documentclass[prl,twocolumn]{revtex4}
\usepackage{graphicx}
\usepackage{amssymb}
\usepackage{amsmath}
\usepackage{enumitem}

\begin{document}

\title{Uniformity of the pseudomagnetic field in strained graphene}

\author{G. J. Verbiest}
\author{S. Brinker}
\author{C. Stampfer}
\affiliation{JARA-FIT and 2nd Institute of Physics, RWTH Aachen University, 52074 Aachen, Germany}

\begin{abstract}
We present a study on the uniformity of the pseudomagnetic field in graphene as a function of the relative orientation between the graphene lattice and straining directions. For this, we strained a regular micron-sized graphene hexagon by deforming it symmetrically by displacing three of its edges. By simulations, we found that the pseudomagnetic field is strongest if the strain is applied perpendicular to the armchair direction of graphene. For a hexagon with a side length of 1 ${\rm \mu}$m, the pseudomagnetic field has a maximum of 1.2 T for an applied strain of 3.5\% and it is uniform (variance $< 1$\%) within a circle with a diameter of $\sim 520$ nm. This diameter is on the order of the typical diameter of the laser spot in a state-of-the-art confocal Raman spectroscopy setup, which suggests that observing the pseudomagnetic field in measurements of shifted magneto-phonon resonance is feasible.
\end{abstract}

\maketitle

\section{Introduction}

Graphene is well-known for its outstanding electronic and mechanical properties as well as their remarkable coupling \cite{Huang,Trif}.
Electronically, graphene behaves as a high mobility semimetal \cite{Katsnelson2}, and mechanically, it is the strongest material known to mankind \cite{Lee}. That graphene is a semimetal constraints the currently reachable on/off ratios for graphene transistors \cite{Xia}.
Therefore, much research is focussed on this electromechanical coupling, as it may be utilized to tailor the electronic properties of graphene.
In theory, it is even possible to get the desired electronic properties completely via so-called strain engineering \cite{Pereira,Low}.
In particular, a large mechanical deformation could induce a bandgap \cite{Hasegawa,Pereira2,Pellegrino,Cocco,Choi}. However, even small deformations, or equivalently, small strain fields, already alter the electronic properties.
These small deformations generate an effective potential for the charge carriers in graphene,
which is similar to the vector potential of a {\it real} magnetic field and is therefore called the {\it pseudovector potential} \cite{Suzuura,Sasaki,Katsnelson,Juan}.
In analogy to a real static magnetic field, a pseudovector potential can generate a {\it pseudomagnetic} field \cite{Masir,Adebpour,Kim}.
This pseudomagnetic field is always zero, unless the applied strain is {\it not uniform}.

A major hurdle in strain engineered graphene is the (experimental) determination of the pseudomagnetic field strength and its uniformity.
Recently, it was shown that the pseudomagnetic field in graphene nanobubbles can be as strong as 300 T \cite{Levy},
which is much stronger than one can achieve with real magnetic fields.
Therefore, strained graphene offers the unique opportunity to study the electronic properties of graphene at extreme (pseudo)magnetic field strengths.
Alternatively, such systems could allow studies to the magneto-phonon resonance in Raman spectroscopy \cite{Neumann} or enable so-called valley-tronics \cite{Rycerz,Guimaraes,Jiang}.
Numerical work of strain fields in graphene is therefore of additional value, as it not only estimates the strength of the pseudomagnetic field, but also shows the uniformity of the generated pseudomagnetic field and its dependence on the lattice orientation with respect to the strain direction.

In this paper, we show calculations of the pseudomagnetic field in a regular hexagon of graphene that is strained simultaneously in three different directions. This is particularly relevant for experiments employing microelectromechanical systems (MEMS), which exactly allow for these type of strain fields \cite{Garza}. As the uniformity of the pseudomagnetic field determines whether it is detectable with a local probe, e.g. a scanneling tunneling microscope cantilever or a laser probe, we specifically study this as a function of the lattice orientation. For this, we consider two hexagonal geometries with side length $L$ of 100 nm and 1 $\mu$m respectively. We show that the pseudomagnetic field is constant within a circle of diameter $d$ that strongly depends on the relative orientation between the graphene lattice and the strain direction. For the hexagonal geometry with $L = 1$ ${\rm \mu}$m, we find a diameter $d$ of around 520 nm. In addition, we show that a special geometry, which was believed to generate a uniform pseudomagnetic field for a specific strain field \cite{Guinea}, actually does not do so. Finally, we show that an uniform pseudomagnetic field can indeed be generated by applying a linearly varying force on two opposing edges of a rectangle \cite{Guinea2}.

This paper is organized as follows. First, we describe the theoretical origin of the pseudomagnetic field. Second, we discuss the methods that we used to calculate the pseudomagnetic field. Then, we show our results before finishing with the experimental implications and the conclusion.

\section{The Pseudomagnetic Field}

\begin{figure}[!t]
	\begin{center}
		\includegraphics[width=86mm]{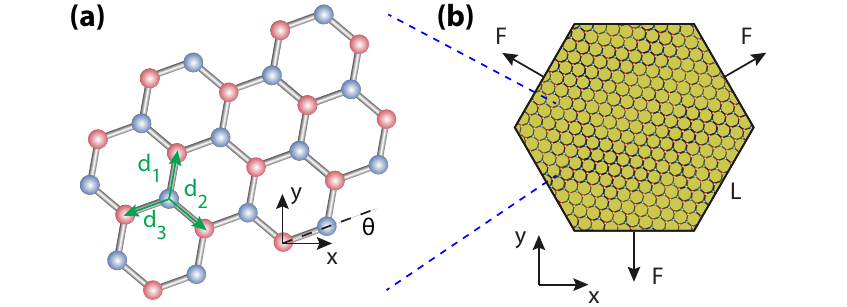}
	\end{center}
	\caption{(a) The lattice structure of graphene, in which $\theta$ indicates the rotation angle between the armchair direction and the x-axis. The red and blue atoms indicate the two different sublattices, and the ${\bf d_n}$ indicates the nearest neighbor vectors. (b) The geometry used in our calculations is a regular hexagon with side length $L$. The force $F$ is applied to the indicated edges.}
	\label{fig1}
\end{figure}

The electronic structure of graphene is described by a tight-binding Hamiltonian which only takes the interaction of the $\pi$-states into account \cite{Katsnelson2}:

\begin{equation}
H = -\sum_{n=1}^3 t_n \left(
\begin{array}{cc}
0 & e^{-i {\bf k \cdot d_n} }\\
e^{i {\bf k \cdot d_n} } & 0
\end{array}
\right),\label{eqH0}
\end{equation}

\noindent
in which ${\bf k}$ is the momentum of the electron, $\{ {\bf d_n} \}$ are the three nearest neighbor vectors (see Fig. \ref{fig1}a), and $\{t_n\}$ are the corresponding nearest neighbor hopping parameters. This Hamiltonian is a 2x2-matrix, as each unit cell of graphene contains two carbon atoms. As Fig. \ref{fig1}a shows, each of these atoms forms a sublattice. The energy spectrum corresponding to Eq. (\ref{eqH0}) contains six specific electron momenta ${\bf k}$, for which the energy is zero. These six points are characterized by only two momenta: ${\bf K}$ and ${\bf K'}$, which is a consequence of the two carbon atoms per unit cell.

The nearest neighbor hopping parameters $\{t_n\}$ change under a deformation of the graphene lattice, as such a deformation changes the nearest neighbor distance. The change in hopping parameters
is described by

\begin{equation}
t_n = t_0 e^{-\beta \left( |{\bf d_n}|/a - 1 \right)},
\end{equation}

\noindent
in which $a$ is the nearest neighbor distance without any deformation, $t_0 \approx 2.8$ eV is the hopping parameter without any deformation, and $\beta \approx 2-3.37$ is the Gr\"{u}neisen parameter \cite{Pereira2,Heeger,Vozmediano,Castro}. Please note that for the calculations below, we follow \cite{Pereira2,Castro} and use a Gr\"{u}neisen parameter of $\beta = 3.37$.

In case of zero strain ($|{\bf d_n}| = a$ $\forall n$), one can make the usual expansion around the {\bf K}-point for low energy electrons (${\bf k} = {\bf K}+ {\boldsymbol\kappa}$, to find a Hamiltonian that is linear in ${\boldsymbol\kappa}$,

\begin{equation}
H = \hbar v_{\rm F} {\boldsymbol \sigma} \cdot {\boldsymbol\kappa}\label{DiracEq},
\end{equation}

\noindent
where $v_{\rm F} = 3 a t_0 / 2\hbar$ is the Fermi velocity, and ${\boldsymbol \sigma} = (\sigma_x,\sigma_y)$ are the Pauli matrices.

In the case of nonzero strain, one finds a similar expression for the Hamiltonian by using $\delta t_n = t_n - t_0$. In fact, one finds the same Hamiltonian as in Eq. (\ref{DiracEq}), but with an additional term that is independent of the momentum ${\boldsymbol\kappa}$:

\begin{equation}
H = \hbar v_{\rm F} {\bf \sigma} \cdot ({\boldsymbol\kappa} - e {\bf A})\label{DiracStrain},
\end{equation}

\noindent
in which $e$ is the electron charge, and ${\bf A} = (A_{\rm x}, A_{\rm y})$ is the so-called pseudovector potential,

\begin{equation}
A_{\rm x} + i A_{\rm y} = - \frac{1}{e v_{\rm F}} \sum_{n=1}^3 \delta t_n e^{-i {\bf K \cdot d_n} }.
\end{equation}

Equation (\ref{DiracStrain}) immediatly explains why ${\bf A}$ is called a pseudovector potential, as the additional term in $H$ mimics the vector potential of a real magnetic field.
Therefore, we define, in analogy to a real magnetic field, a pseudomagnetic field:

\begin{equation}
B = \frac{\partial A_{\rm y}}{\partial x} - \frac{\partial A_{\rm x}}{\partial y}.
\end{equation}

However, there is one remarkable difference: $B$ changes sign, if one expands the Hamiltonian around the ${\bf K'}$-point instead of the ${\bf K}$-point. Consequently, the electrons in the two valleys feel an opposite magnetic field \cite{Kim}. This property is important for valley-tronics since it may allow to make valley filters \cite{Rycerz,Guimaraes,Jiang}.

\section{Rotation of the Lattice relative to the Strain Direction}

It is a convention to define an angle $\theta$ as the angle between the x-axis and the so-called armchair direction of graphene (see Fig. \ref{fig1}a).
If $\theta = 0$ and the displacements $|{\bf u}|$ are much smaller than $a$, it is well known that the pseudovector potential can be written in terms of the deformation tensor $u_{ij}$ [$i,j = (x,y)$] \cite{Suzuura,Manes},

\begin{align}
A_{\rm x} &= \frac{\hbar \beta}{2 a e} (u_{\rm xx} - u_{\rm yy})\label{AxpsDeformation}\\
A_{\rm y} &= \frac{\hbar \beta}{2 a e} ( - 2u_{\rm xy}),\label{AypsDeformation}
\end{align}

\noindent
in which $u_{ij} = \frac{1}{2}\left( \frac{\partial u_i}{\partial j} + \frac{\partial u_j}{\partial i} \right)$.

We now rotate the lattice with respect to the x-axis by an angle $\theta$ (see Fig. \ref{fig1}a) using the standard rotation matrix $R$ in two dimensions:
\begin{equation}
R = \left(
\begin{array}{cc}
\cos\theta & -\sin\theta\\
\sin\theta & \cos\theta
\end{array}
\right).
\end{equation}
\ \\
\noindent
We note that ${\bf K} \cdot {\bf d_n}$ is constant under lattice rotations, which implies that the usual derivation of Eqs. (\ref{AxpsDeformation}) and (\ref{AypsDeformation}) still holds.
Therefore, we only need to find the deformation tensor for the rotated lattice \cite{Guinea}, before we rotate the pseudovector potential {\bf A}:
\begin{eqnarray}
\left( \begin{array}{cc} u_{x'x'} & u_{x'y'} \\ u_{x'y'} & u_{y'y'} \end{array} \right)
&=& \mathcal{R} \left( \begin{array}{cc} u_{xx} & u_{xy} \\ u_{xy} & u_{yy} \end{array} \right) \mathcal{R}^T \label{DefRot}
\end{eqnarray}
\noindent
and
\begin{eqnarray}
\left( \begin{array}{c} A_x \\ A_y \end{array} \right) &=& \left( \begin{array}{cc} \cos\theta & \sin\theta \\ -\sin\theta & \cos\theta\end{array} \right) \left( \begin{array}{c} A_{x'} \\ A_{y'} \end{array} \right). \label{ARot}
\end{eqnarray}
Substituting Eqs. (\ref{DefRot}) and (\ref{ARot}) into Eqs. (\ref{AxpsDeformation}) and (\ref{AypsDeformation}), we find an expression for the pseudovector potential for arbitrary lattice rotations \cite{Guinea}:
\begin{eqnarray}
A_x =& \frac{\hbar \beta}{2 a e} \left( (u_{xx}-u_{yy}) \cos(3\phi)-2 u_{xy} \sin(3\phi) \right)\label{AxRotF} \\
A_y =& -\frac{\hbar \beta}{2 a e} \left( (u_{xx}-u_{yy}) \sin(3\phi) + 2 u_{xy} \cos(3\phi) \right).\label{AyRotF}
\end{eqnarray}
Considering the pseudomagnetic field, one sees that it has a rotational symmetry of $120^{\circ}$. A rotation of the graphene lattice by an angle of $120^{\circ}$ gives the same pseudomagnetic field and a rotation of $60^{\circ}$ effectively changes the sign. This implies that the pseudomagnetic field is zero for an angle of $30^{\circ}$ degrees.

\section{Methods}

We define a hexagonal geometry with side length $L$, as shown in Fig. \ref{fig1}b, of which the orientation is always kept constant. It is only the lattice orientation of the graphene that is rotated with respect to the x-axis.
In order to calculate the pseudomagnetic field we make use of a 2-step approach.

In the first step we calculate the deformation of the geometry in COMSOL \cite{COMSOL}, for which we consider only in-plane strain as well as only the {\it linear} contributions to elasticity theory. For this calculation, we assumed that graphene has a thickness of 3.35 \AA, a Young's modulus of 1.02 TPa, and a Poisson's ration of 0.165 \cite{Lee,Blakslee}. The strain is applied via a fixed displacement on the three edges that are indicated with arrows in Fig. \ref{fig1}b. From the calculated deformation field, we export the displacement vectors at the position of the atoms for the second step, in which we calculate both the pseudovector potential and the pseudomagnetic field.

\section{Results}

\begin{figure}[!b]
	\begin{center}
		\includegraphics[width=86mm]{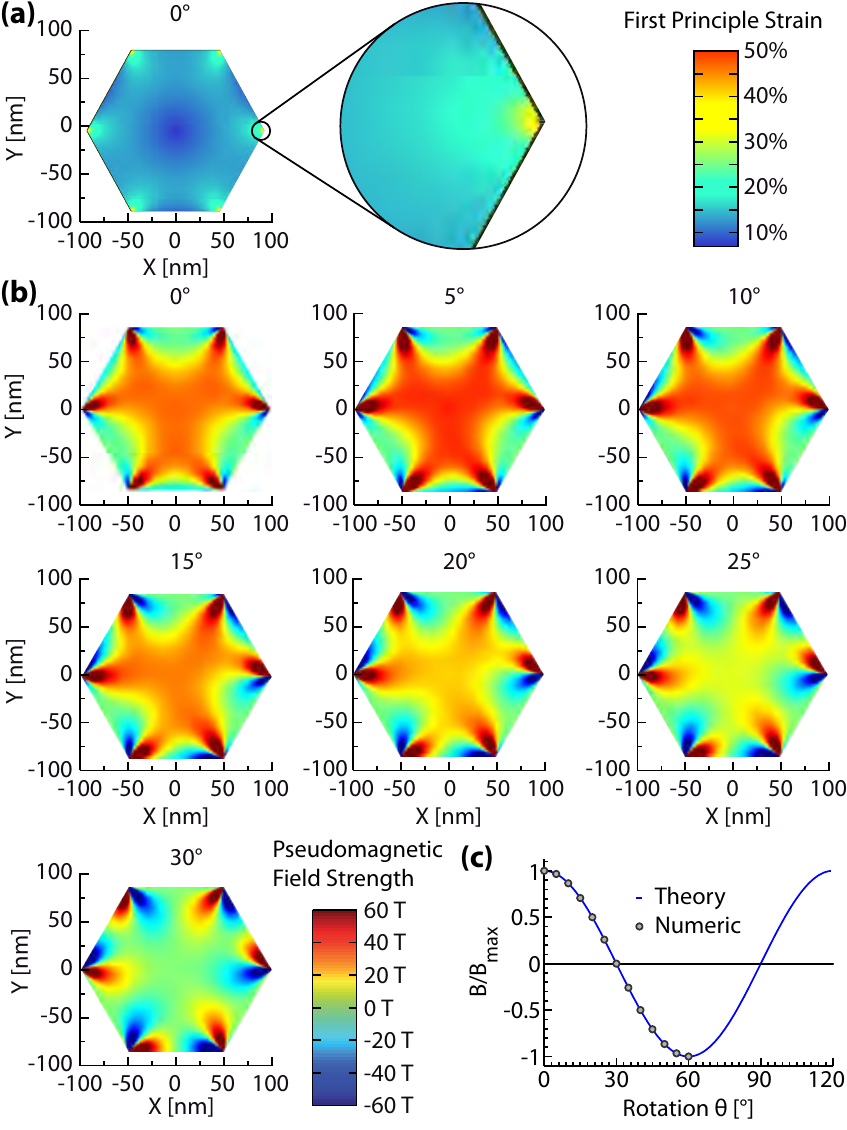}
	\end{center}
	\caption{(a) The strain in the geometry for zero rotation angle $\theta$. There are three edges with zero strain and three with $~ 15\%$ of strain. The corners of the geometry exhibit the largest amount of strain (up to $~ 50\%$), whereas the strain reduces to 0 \% in the center. (b) The pseudomagnetic field as a function of the rotation angle $\theta$. At $\theta = 30^{\circ}$ the pseudomagnetic field disappears in the center of the geometry, as the atomic displacement is parallel to the nearest neighbor vectors. (c) The pseudomagnetic field in the center of the geometry is in excellent agreement with that from Eq. (\ref{AxRotF}) and (\ref{AyRotF}). Please note that the pseudomagnetic field is normalized with respect to the field at $\theta = 0^{\circ}$.}
	\label{fig2}
\end{figure}

Let us start with a hexagonal geometry of $L = 100$ nm. Figure \ref{fig2}a shows the strain obtained by applying a fixed displacement of 10 nm to the three edges for zero rotation angle ($\theta = 0$). As three sides have a fixed displacement, the side length $L$ of those particular sides is fixed to that of the undeformed geometry. In contrast, the side lengths of the other sides {\it are} allowed to change. This exactly leads to the strain pattern that we observe: zero strain along three edges at which we apply the fixed displacement and a finite strain (15\%) along the other three edges. For the same reason, the strain is highest at the six corners of the geometry (up to $~ 50\%$). Please note that this 50\% of strain will not be reached in an experimental situation, as the corners will not be so sharp and the boundary condition will not be so rigid. In the center of the geometry, the strain reduces to 0\%. This is due to the symmetry of the applied fixed displacements, as the (vector) sum of them is zero.

The panels in Fig. \ref{fig2}b show the pseudomagnetic field for rotation angles $\theta$ varying from 0 to 30 degrees. A positive (negative) pseudomagnetic field is indicated in red (blue). We observe two contributions to the total pseudomagnetic field: the one at the six corners of the hexagon and the one from the center of the geometry. The pseudomagnetic field at the corners does not change when rotating the graphene lattice. It remains constant with values around $+60$ T and $-60$ T. The origin of this pseudomagnetic field is our boundary condition (see above). As a consequence, we get a large relative displacement at the six corners of the geometry, which results in the observed pseudomagnetic field.

In contrast, the pseudomagnetic field in the center of the geometry does depend on the rotation angle $\theta$. The maximum pseudomagnetic field $B_{\rm max}$ in the center of the geometry is around $40$ T for $\theta = 0^{\circ}$ and decreases rapidly when increasing the rotation angle $\theta$. At a rotation angle of $\theta = 30^{\circ}$, the pseudomagnetic field in the center of the geometry is even {\it zero}, as the atomic displacement is parallel to the nearest neighbor vectors. Figure \ref{fig2}c shows the pseudomagmetic field in the center of the geometry normalized by $B_{max}$ as a function of the rotation angle $\theta$ as well as the normalized pseudomagnetic field that one obtains from Eqs. (\ref{AxRotF}) and (\ref{AyRotF}). As Fig. \ref{fig2}c shows, our numerical calculation is in perfect agreement with the theoretical prediction.

\begin{figure}[!b]
	\begin{center}
		\includegraphics[width=86mm]{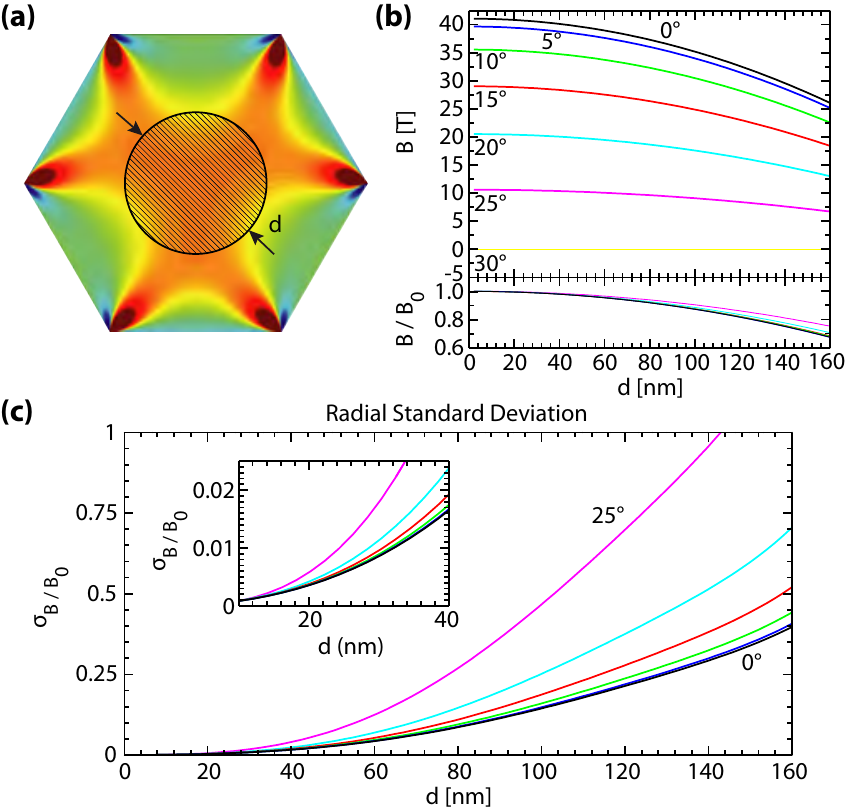}
	\end{center}
	\caption{(a) The average pseudomagnetic field and its corresponding normalized standard deviation are calculated within a circle of diameter $d$ that is centered in the geometry. (b) The upper panel shows the pseudomagnetic field as a function of the diameter $d$ for different rotation angles. The bottom panel shows that all these curves fall almost on top of each other if one normalizes each curve with respect to the value of the pseudomagnetic field in the center. (c) The standard deviation of the pseudomagnetic field as a function of the diameter $d$ for the rotation angles shown in (b). The generated pseudomagnetic field is uniform within $d = 34$ nm for $\theta = 0^{\circ}$. The inset shows a zoom for small $d$.}
	\label{fig3}
\end{figure}

The uniformity of the pseudomagnetic field is of crucial importance for the possible utilization of it in experiments and devices. To quantify the uniformity of the pseudomagnetic field $B$, we calculate the average value of $B$ in a circle of diameter $d$ that is centered in the geometry (see Fig.~\ref{fig3}a). The result is shown in the upper panel of Fig.~\ref{fig3}b for various rotation angles $\theta$. In order to compare the uniformity of the pseudomagnetic field at different rotation angles $\theta$, we normalize each curve with respect to the pseudomagnetic field $B_0$ at $d = 2$ nm (see the lower panel of Fig. \ref{fig3}b). Please note that the curves for different $\theta$ are very similar and only deviate for large $d$. Finally, Fig. \ref{fig3}c shows the standard deviation of the normalized pseudomagnetic field $\sigma_{B/B_0}$ as a function of $d$. To characterize the homogeneous area of the pseudomagnetic field one has to set an upper limit on the standard deviation. We choose a standard deviation smaller than 1\% of the maximum pseudomagnetic field in the center. By using this definition, we obtain a diameter of uniformity of around $34$ nm for $\theta = 0$.

\begin{figure}[!b]
	\begin{center}
		\includegraphics[width=86mm]{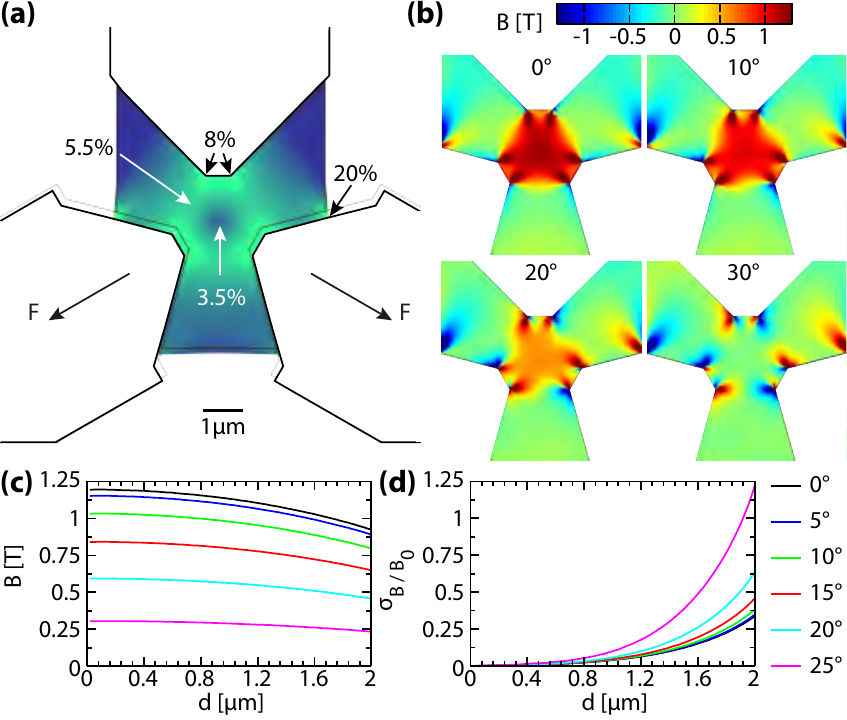}
	\end{center}
	\caption{(a) The computed strain field in a graphene sheet that can be reached with a state-of-the-art MEMS device shows values ($\sim 20\%$) close to the maximum strain that graphene can withstand. The force $F$ is applied to the indicated edges. (b) The corresponding pseudomagnetic field for various rotation angles $\theta$. (c) shows the pseudomagnetic field as a function of $d$ and (d) shows the corresponding standard deviation. In this case, the pseudomagnetic field is uniform within $d = 520$ nm.}
	\label{fig4}
\end{figure}

Figure \ref{fig4}a shows an overview of a feasible MEMS actuator that allows to experimentally strain graphene sheets. The arms of the MEMS actuator are placed under an angle of 120 degrees and are connected with a rectangular sheet of graphene, which has a width of 5 ${\rm \mu}$m and a length of 7 ${\rm \mu}$m. This configuration generates a force field, which has the same threefold symmetry as for the hexagonal structure above (see Fig.~\ref{fig1}c). However, the effective side length is now $\sim 1$ ${\rm \mu}$m. The strain in the graphene sheet resulting from the same fixed displacement as before is shown in Fig. \ref{fig4}a. Although the strain in the center region of the geometry is on average 3.5\%, the maximum strain value is 8\% on the corners of the hexagon and 5.5\% on the sides. The maximum strain in the entire geometry occurs at the positions where the graphene/MEMS device boundary has a corner. At these locations, the strain is even around 20\%, which is the amount of strain at which the graphene sheet is supposed to start to rupture \cite{Lee,Kim2}. Nevertheless, it is expected that such high strain values will not be reached in an experiment.

The pseudomagnetic field for different rotation angles of the lattice with respect to the strain directions is calculated and depicted in Fig. \ref{fig4}b. The pseudomagnetic field has the same characteristics as the pseudomagnetic field of the strained hexagon in Fig. \ref{fig2}. The strength of the pseudomagnetic field is highest in each corner of the hexagon and the field in the center is strongest for $\theta = 0$. Figure \ref{fig4}c shows the strength of the pseudomagnetic field as a function of the diameter $d$ of a circle that is placed identically to that in Fig. \ref{fig3}a. For $\theta = 0$, the pseudomagnetic field is approximately 1.2 T in the center. If we compare this with the maximum pseudomagnetic field in the hexagon with side length $L = 100$ nm, we see that it is approximately 30 times weaker. This is due to the following two effects. Firstly, by increasing the side length $L$ by a factor of 10, the gradient in the pseudovector potentials is 10 times smaller, which results in a 10 times smaller pseudomagnetic field. Secondly, the applied strain is now approximately $\sim 3.5$\%, which is approximately 3 times smaller than the 10\% that we had before. Combining both, explains the $\sim 30$ times weaker pseudomagnetic field.

If we now consider the uniformity of the pseudomagnetic field in such MEMS devices, as shown in Fig. \ref{fig4}d, we obtain a diameter of uniformity of around $520$ nm for $\theta = 0^{\circ}$. If we would choose a standard deviation smaller than 5\% of the maximum pseudomagnetic field in the center of the geometry, the diameter of uniformity increases to $\sim 1100$ nm for $\theta = 0^{\circ}$. As both values are comparable to the typical laser spot size in state-of-the-art confocal Raman spectroscopy experiments, it may be possible to quantify the pseudomagnetic field strength in an experiment as it induces a shift in the MPR.

\begin{figure}[!b]
	\begin{center}
		\includegraphics[width=86mm]{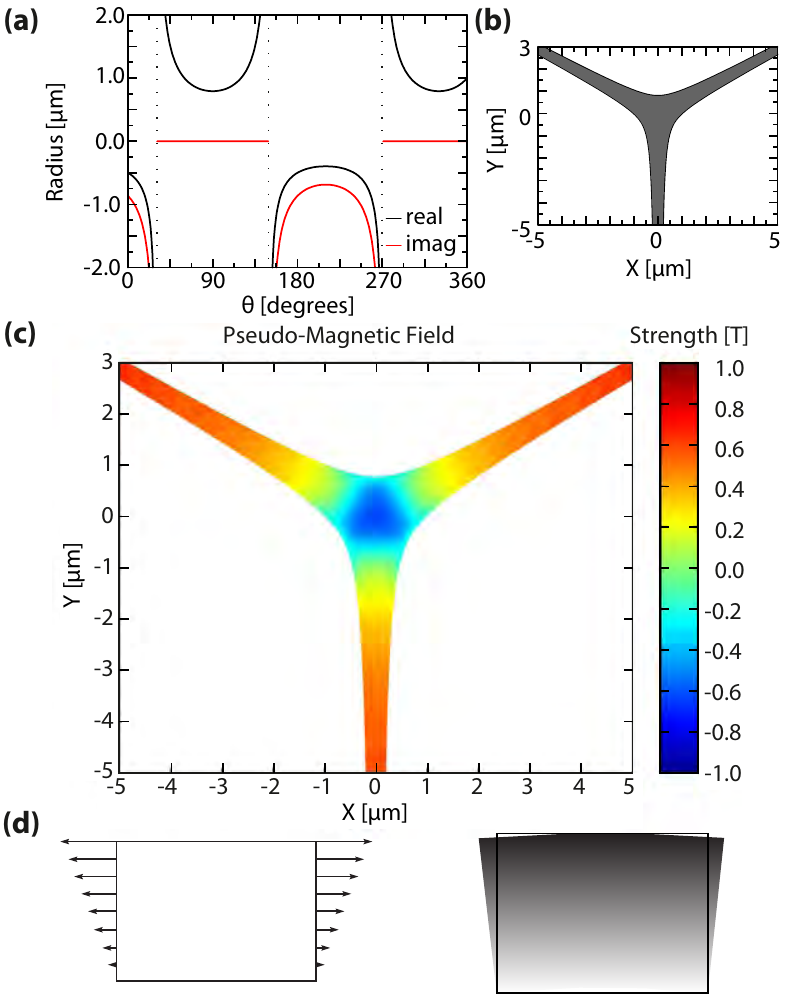}
	\end{center}
	\caption{The solution for the radius $r$ for $c=1$ $\mu$m as a function of the angle $\theta$ ({\bf a}) and its corresponding geometry for $|r|$ ({\bf b}). In contrast to what was thought, the resulting pseudomagnetic field is not uniform ({\bf c}). (d) The linearly varying forces (left panel) on two opposing boundaries of a rectangle that generate an uniform pseudomagnetic field. The right panel shows the corresponding deformation of the geometry.}
	\label{fig5}
\end{figure}

Apart from the experimentally accessible geometries that we discussed so far, certain geometries with corresponding force fields have been suggested that should give rise to a completely uniform pseudomagnetic field. In particular, Guinea {\it et al.} \cite{Guinea} give two of such (analytic) geometries with corresponding force fields. Using our approach, we confirmed the uniform pseudomagnetic field in the circular disk for the displacement field given in Ref. \cite{Guinea}. Let us next focus on the second geometry given in Ref. \cite{Guinea}. This geometry should give an uniform pseudomagnetic field when only perpendicular forces are applied to its boundary. The analytic expression for this geometry in polar coordinates is derived under the assumption of pure shear stresses (see the Supplementary Information of \cite{Guinea}):

\begin{equation}
r(\phi) = \frac{c}{\left[ (\cos\phi/2 \mp \sin\phi/2)(\pm 1 + 2\sin\phi)\right]^{2/3}}.\label{Eq1}
\end{equation}

\noindent
In this equation, $r$ is the radius, $\phi$ is the polar angle, and $c$ is a constant. The $\pm$ sign indicates two equivalent geometries that are rotated by 60 degrees with respect to each other. It does not matter which sign we take, the radius always becomes imaginary for specific angles. For simplicity, we choose the + sign for our geometry. Figure~\ref{fig5}a shows both the real part (black) and imaginary part (red) of the radius $r$ as a function of the angle $\phi$ for $c = 1$ $\mu$m. Please realize that $r(\phi)$ is imaginary and, therefore, physically {\it not} allowed. We can obtain the geometry shown in \cite{Guinea} by taking the absolute value of $r(\phi)$ (see Fig. \ref{fig5}b). However, if we calculate the pseudomagnetic field for this given geometry and force field using our numerical approach, we observe that it is not uniform (see Fig. \ref{fig5}c). However, in the arms of this geometry, the pseudomagnetic field looks almost uniform. Based on this, it can be shown that a rectangle with linearly varying forces perpendicular to two opposing boundaries (see left panel in Fig. \ref{fig5}d) does give rise to an uniform pseudomagnetic field \cite{Guinea2}. The corresponding deformation of the geometry is shown in the right panel of Fig. \ref{fig5}d.

\section{Experimental Implications}

A particular experiment, in which the strength of a tunable pseudomagnetic field may be directly probed, is the shift of the magneto-phonon resonance (MPR) in Raman spectroscopy \cite{Goerbig,Faugeras,Goler,Faugeras2}. In order to exclude other effects, the pseudomagnetic field must be uniform within the laser spot size in the confocal Raman spectroscopy experiment, which is exactly what we showed and discussed in this paper. By following the recent experiment of Neumann {\it et al.} \cite{Neumann}, where MPRs at low magnetic fields have been observed, we can estimate the experimental requirements for observing MPR shifts thanks to a tunable pseudomagnetic field. The most prominant MPR below 20 T is the so-called $T_1$ transistion at around 3.7 T, which results in an well resolved peak of the Raman G-line width as function of magnetic field. This resonance peak has an experimentally observed peak line-width (FWHM) of around 700 mT. Assuming that we need for unambigously proving the presence of a pseudomagnetic field  at least a peak shift of half the line-width of this magneto-phonon resonance. This results in a required pseudomagnetic field of around 350~mT, which corresponds to a strain value of 1\% in our proposed device. These values are by a factor 3.5 below the values discussed above (see Fig. \ref{fig4}a) allowing for a non-perfect lattice alignment and making this approach promising for investigating tunable pseudomagentic fields.

\section{Conclusion}

We studied the strength and uniformity of the pseudomagnetic field in a hexgonal sheet of graphene as a function of the rotation angle $\theta$ between the armchair direction of graphene and the strain direction. We showed that the relative orientation of the graphene lattice with respect to the strain direction is extremely important, as the pseudomagnetic field even disappears for specific angles. The pseudomagnetic field is strongest when the strain is parallel to the armchair direction in the graphene. To characterize the homogeneous area of the pseudomagnetic field, we set an upper limit on the standard deviation in the pseudomagnetic field of 1\%. Using this definition, the pseudomagnetic field is constant within a diameter of $\sim 520$ nm for a hexagonal graphene sheet with a side length of 1 ${\rm \mu}$m. Our results show that the pseudomagnetic field is detectable with a local probe such as a scanneling tunneling microscope cantilever but even also with a typical confocal laser probe. One may even think of observing a shift in the magneto-phonon resonance due to the strain-induced pseudomagnetic field. In addition, tunable pseudomagnetic fields are important for valley-tronics since they may allow to make valley filters.

\section{Acknowledgement}

We acknowledge funding from the ERC (GA-Nr. 280140) and the EU Flagship-Graphene (contract no. NECT-ICT-604391).

\end{document}